\documentclass[amsmath,amssymb,superscriptaddress,prd,nofootinbib,twocolumn]{revtex4-2}
\usepackage{graphicx}
\usepackage{amsmath}
\usepackage{amssymb}
\usepackage{amsfonts}
\usepackage{color}
\usepackage{dsfont}
\usepackage{tikz}
\usepackage{ragged2e}
\usetikzlibrary{calc,decorations.markings}
\usepackage[colorlinks=true,linkcolor=blue,citecolor=blue,urlcolor=blue]{hyperref}
\usepackage{tipa}
\usepackage{physics}
\usepackage{soul}
\usepackage{float}
\usepackage{subfig}
\newcommand{\kk}{\boldsymbol{k}}
\newcommand{\nn}{\nonumber}
\newcommand{\xx}{\mathbf{x}}
\newcommand{\x}{\mathsf{x}}

\usetikzlibrary{decorations.pathmorphing, patterns,shapes}

\begin{document}

\title{Uniformly accelerated Brownian oscillator in (2+1)D: temperature-dependent dissipation and frequency shift}
\date{\today}

\author{Dimitris Moustos}\email{dmoustos@upatras.gr}\affiliation{Department of Physics, University of Patras, 26504 Patras, Greece}

\begin{abstract}

We consider an Unruh-DeWitt detector modeled as a harmonic oscillator that is coupled to a massless quantum scalar field in the (2+1)-dimensional Minkowski spacetime. We treat the detector as an open quantum system and employ a quantum Langevin equation to describe its time evolution, with the field, which is characterized by a frequency-independent spectral density, acting as a stochastic force. We investigate a point-like detector moving with constant acceleration through the Minkowski vacuum and an inertial one immersed in a  thermal reservoir at the Unruh temperature, exploring the implications of the well-known non-equivalence between the two cases on their dynamics. We find that both the accelerated detector's dissipation rate and the shift of its frequency caused by the coupling to the field bath depend on the acceleration temperature. 
Interestingly enough this is not only in contrast to the case of inertial motion in a heat bath but also to any analogous quantum  Brownian motion model in open systems, where dissipation and frequency shifts are not known to exhibit temperature dependencies. Nonetheless, we show that the fluctuating-dissipation theorem still holds for the detector-field system and in the weak-coupling limit an accelerated detector is driven at late times to a
thermal equilibrium state at the Unruh temperature.
\end{abstract}

\maketitle

\section{Introduction}

The Unruh effect \cite{Fulling,Davies,Unruh} asserts that observers moving with a constant acceleration of magnitude $a$ in Minkowski spacetime perceive the Minkowski vacuum as a thermal state at a temperature proportional to their acceleration, known as the Unruh temperature $T_U= \hbar a/(2\pi ck_B)$, where $\hbar$ is the reduced Planck constant, $c$ the
speed of light and $k_B$ the Boltzmann’s constant. It is a direct manifestation of the observer dependence of the notion of the vacuum and hence of the particle content in quantum field theory \cite{birrell,wald:book}.

Typical considerations of the Unruh effect employ the concept of the Unruh-DeWitt (UDW) particle detector \cite{Unruh,DeWitt}: a point-like two-level system that interacts
locally with a quantum field through a monopole coupling while moving
along a worldline in Minkowski spacetime. The excitation rate of a uniformly accelerated detector, initially prepared in its ground state, is then evaluated to leading order in time-dependent perturbation theory \cite{birrell} and is found to follow a Planck distribution at the Unruh temperature.

Nonetheless, the identification of the Unruh effect through the Planckian form of the transition rate has proven to be misleading. This is mainly because the particle detector approach is highly model-dependent. Different types of detectors and couplings to fields result in different detector responses to the vacuum fluctuations of a field \cite{Earman,Unruh:review}. One striking example is the dependence of the excitation rate of an accelerated detector in a scalar field background on the number of the dimensions of the underlying spacetime, with the  Bose-Einstein distribution observed for even dimensions being replaced by a Fermi-Dirac one in odd dimensions \cite{TakagiInv,Takagi}. 

On the other hand, treating the detector as an open quantum system \cite{breuer} with the field playing the role of the environment it can be shown that in the long-time limit an accelerated detector reaches a Gibbs state at the Unruh temperature regardless of the details of the  interaction or the intermediate dynamics (see, e.g., \cite{Benatti,DeBievre:Merkli,DMCA,DM,DMBJA,Kaplanek,Garay}). The late time behavior of the detector provides a more rigorous and universal way to interpret the Unruh effect and its thermal character \cite{DMCA,DM}.

In the present work, we model the detector as  a harmonic oscillator rather than a qubit. The oscillator detector model is equivalent to a quantum Brownian motion (QBM) model \cite{Feynman,CALDEIRA,GrabertQBM,Hu:Paz:Zhang} of an oscillator coupled to a bath of an infinite number of non-interacting harmonic oscillators. We study the response of an accelerated detector interacting with a massless quantum scalar  field in its vacuum state in the (2+1)-dimensional Minkowski spacetime. Working in the framework of open systems we use a quantum Langevin equation \cite{WeissQDS,Ford:OConnell} to describe its time evolution with the effects of the field bath being incorporated in the Pauli-Jordan and the Hadamard functions of the field. Our aim is to explore the consequences of the statistics reported in \cite{TakagiInv,Takagi} as well as the emergence of thermality in the evolution of a QBM detector model in a (2+1) dimensional spacetime background. 

We consider the cases of a point-like detector moving with constant acceleration through the Minkowski vacuum and a static one immersed in a heat bath at the Unruh temperature. Although the accelerated oscillator behaves in exactly the same way as the static one in (1+1) and (3+1) spacetime dimensions, this equivalence ceases to hold when considering the (2+1)-dimensional case \cite{Takagi}. We ask then the following question:
\emph{What this  non-equivalence in (2+1) dimensions implies for the dynamics of an accelerated Brownian oscillator?}

 We find that, as a result of this contradiction, both  the accelerated detector's dissipation rate and the Lamb shift depend on the acceleration temperature. This is in contrast to the case of inertial motion in heat bath and most importantly--to our knowledge--to any conventional QBM model, where dissipation and frequency shifts are not known to exhibit temperature dependencies. Finally, we show that despite the aforementioned discrepancy the fluctuating-dissipation theorem still holds and in the weak-coupling limit an accelerated detector asymptotically reaches at late times a thermal state at the Unruh temperature.

Throughout the paper we denote  spacetime vectors with sans-serif characters $(\x)$, while spatial vectors are represented by boldface letters $(\xx)$. We use the signature $(+--\dots-)$ for the Minkowski metric. Unless otherwise specified we hereafter  set  $\hbar=c=k_B=1$.

\section{The quantum Langevin equation and its solution}

We model an UDW detector as a harmonic oscillator with unit mass and bare frequency $\Omega$, whose position operator $\hat{x}$ is linearly coupled to a massless quantum scalar field through the Hamiltonian \cite{Unruh:Zurek,Hu:Louko}
\begin{align}\label{UDW:Ham}
    H_I=\lambda\hat{x}\otimes\hat\Phi(\xx),
\end{align}
where $\lambda$ is the coupling constant and $\hat\Phi(\xx)$ is the pullback of the field to the detector's position $\xx$.

The oscillator  detector  model described by the interaction Hamiltonian \eqref{UDW:Ham} is a special case of the Caldeira-Leggett \cite{CALDEIRA} model of QBM, which has been extensively employed in the theory of open quantum systems to investigate phenomena such as dissipation and decoherence that appear in a Brownian particle when it is coupled to a bath comprised by an infinite number of non-interacting harmonic oscillators. The bath is usually assumed to be initially in a (Gaussian) thermal state with a finite temperature. In the case considered here, the quantum field acts as the environment.

In the Heisenberg picture, the time evolution of the detector's position operator is given by the quantum Langevin equation \cite{Sciama,Hu:Matacz}
\begin{align}\label{Langevin:eq}
    \ddot{\hat{x}}(\tau)+\Omega^2\hat{x}(\tau)+2\int_0^\tau ds \chi(\tau-s)\hat{x}(s)=\hat{\varphi}(\tau),
\end{align}
where $\hat\varphi(\tau):=\lambda\hat\Phi(\x(\tau))$ plays the role of a fluctuating force that obeys Gaussian statistics with $\expval{\hat \varphi(\tau)}=0$ 
 and
\begin{align}
    \chi(\tau,\tau'):=-\frac{i}{2}\expval{\big[\hat{\varphi}(\tau),\hat{\varphi}(\tau')\big]}
\end{align}
 is the \textit{dissipation kernel}. The detector's worldine $\x(\tau)=(t(\tau),\xx(\tau))$ is  parametrized by its proper time $\tau$. 

The solution of the integro-diffrential equation \eqref{Langevin:eq} is
\begin{align}\label{X:solution}
   \hat{x}(\tau)=\dot{G}(\tau)\hat{x}(0)+G(\tau)\hat{p}(0)+\int_0^\tau ds\, G(\tau-s)\hat{\varphi}(s),
\end{align}
where 
$G(\tau)$ is the solution of the homogeneous part of Eq. \eqref{Langevin:eq} with initial conditions $G(0)=0$ and $\dot{G}(0)=1$. The homogeneous solution can be  expressed as an inverse Laplace transform through the Bromwich integral
\begin{align}\label{homo:lapl}
    G(\tau)=\frac{1}{2\pi i}\int_{\alpha-i\infty}^{\alpha+i\infty}\frac{e^{z\tau}}{z^2+\Omega^2+2\widehat{\chi}(z)}dz,
\end{align}
where $\widehat{\chi}(z)$ denotes the Laplace transform of the dissipation kernel and $\alpha$ is a real constant that is larger than the real part of all the
singularities of the integrand.


As the Hamiltonian of the oscillator detector is quadratic to positions and momenta, its state is fully described by its first moments $\expval{R_n}$ and the covariance matrix $\boldsymbol{\sigma}$ of its second moments \cite{Ferraro2005GaussianSI}
\begin{equation}
    \sigma_{nm}:=\frac{1}{2}\expval{\{\hat{R}_n,\hat{R}_m\}}-\expval{\hat{R}_n}\expval{\hat{R}_m},
\end{equation}
where $\hat R_n$ is an element of the vector $\hat{\mathbf{R}}=(\hat{x},\hat{p})^\top$, $\{\cdot,\cdot\}$ stands for the anticommutator and $\expval{\cdot}$ is the average taken over the initial state.

In terms of the solution \eqref{X:solution} the covariance matrix elements are given by 
\begin{align}\label{sigma:xx}
    \sigma_{11}(\tau)&=\expval{\hat{x}^2(\tau)}\nn\\&=\int_{0}^{\tau}ds\int_{0}^{\tau}ds'G(\tau-s)G(\tau-s')\nu(s,s'),
\end{align}
\begin{align}
    \sigma_{22}(\tau)&=\expval{\hat{p}^2(\tau)}\nn\\&=\int_{0}^{\tau}ds\int_{0}^{\tau}ds' \dot{G}(\tau-s) \dot{G}(\tau-s')\nu(s,s'),
\end{align}
\begin{align}
    \sigma_{12}(\tau)&=\frac{1}{2}\expval{\hat{x}(\tau)\hat{p}(\tau)+\hat{p}(\tau)\hat{x}(\tau)}\nn\\&=\int_{0}^{\tau}ds\int_{0}^{\tau}ds'G(\tau-s)\dot{G}(\tau-s')\nu(s,s'),
\end{align}
where 
\begin{align}
    \nu(\tau,\tau'):=\frac{1}{2}\expval{\{\hat{\varphi}(\tau),\hat{\varphi}(\tau')\}}
\end{align}
is the \textit{noise kernel}. We have assumed that the detector and the field are initially prepared in an uncorrelated state, i.e., $\hat{\rho}(0)=\hat{\rho}_D(0)\otimes\hat{\rho}_{\Phi}(0)$, and that $\expval{\hat x(0)}=\expval{\hat p(0)}=0$. Note that since $\expval{\hat \varphi(\tau)}=0$ the first moments vanish. 

We notice that the effects of the field bath are represented in the dissipation and noise kernels. They can be identified with the Pauli-Jordan and the Hadamard functions of the field  respectively \cite{birrell} and together they constitute the Wightman two-point correlation function of the field
\begin{align}\label{infkern}
    \mathcal{W}(\tau,\tau')&=\expval{\hat\varphi(\tau)\hat\varphi(\tau')}\nn\\&\equiv \nu(\tau,\tau')+i\chi(\tau,\tau')
\end{align}
evaluated along the detector's trajectory. We note that when the state of the field is stationary and the detector follows a stationary spacetime trajectory \cite{Letaw}--special cases of which are the inertial and the linear with constant acceleration--the Wightman function depends only on the proper time deference $\tau-\tau'$ between the two points on the detector's worldline and it can be expressed as $\mathcal{W}(\tau,\tau')=\mathcal{W}(\tau-\tau')$.

\section{Uniformly accelerated oscillator detector}\label{Acc:sec}

We consider an oscillator detector that is uniformly accelerated in the $z$ direction of the (2+1)-dimensional Minkowski spacetime. In order to describe the detector's motion it is convenient to introduce the Rindler coordinates  $(\eta,\xi,x_{\bot})$ \cite{Rindler,Rindler:Kruskal}, where $x_{\bot}$ denotes the  spatial coordinate transverse to the direction of the acceleration. They are related to the Minkowski coordinates $(t,z,x_{\bot})$ through the transformation
\begin{equation}\label{acc:traj}
t=a^{-1}e^{a\xi}\sinh(a\eta),\quad z=a^{-1}e^{a\xi}\cosh(a\eta),
\end{equation} under which the Minkowski line element takes the form
\begin{align}
    ds^2=e^{2a\xi}(d\eta^2-d\xi^2)-dx_\bot^2, 
\end{align}
where $\eta$ and $\xi$ take values in the range $-\infty<\eta,\xi<\infty$ and $a$ is a positive constant. They cover the spacetime region with $|t|< z$, known as the right Rindler wedge. An observer moving with constant proper acceleration $a$ can then be described as a static one that follows the trajectory with $\xi=0$ and proper time $\eta$.

In the Rindler coordinates the Klein-Gordon equation $\sqrt{-g}^{-1}\partial_{\mu}\left(\sqrt{-g}g^{\mu\nu}\partial_\nu\right)\hat\Phi=0$ satisfied by a massless scalar field takes the form
\begin{align}
    \frac{\partial^2}{\partial \eta^2}\hat{\Phi}(\eta,\xi,x_{\bot})=\bigg( \frac{\partial^2 }{\partial \xi^2}+e^{2a\xi}\frac{\partial^2}{\partial x_{\bot}^2}\bigg)\hat{\Phi}(\eta,\xi,x_{\bot}),
\end{align}
and the  field operator can be written in terms of the creation $\hat{a}_{\omega k_{\bot}}^{R\dag}$ and annihilation $\hat{a}_{\omega k_{\bot}}^R$  operators of the Rindler modes as \cite{Unruh:review,Takagi}
\begin{equation}
\hat{\Phi}(\eta,\xi,x_{\bot})=\int_0^\infty d\omega\int_{-\infty}^{\infty} dk_{\bot}\left(\upsilon_{\omega k_{\bot}}^R\hat{a}_{\omega k_{\bot}}^R+ \text{H.c.} \right),    
\end{equation}
where
\begin{equation}
  \upsilon_{\omega k_{\bot}}^R =\sqrt{\frac{\sinh(\pi\omega/a)}{2\pi^3a}} K_{i\omega/a}\left(\frac{|k_{\bot}|}{ae^{-a\xi}}\right)e^{-i(\omega\eta-k_{\bot}x_{\bot})},\end{equation}
 is the positive frequency mode function, $k_{\bot}$ the transverse momentum and $K_{\nu}(x)$ is the modified Bessel function of second kind \cite{NIST}.  The creation and annihilation operators satisfy the commutation relations
\begin{align}
    [\hat{a}_{\omega k_{\bot}}^R,\hat{a}_{\omega'k'_{\bot}}^{R\dagger}]&=\delta(\omega-\omega')\delta(k_{\bot}-k'_{\bot}),\nn\\ [\hat{a}_{\omega k_{\bot}}^R,\hat{a}_{\omega'k'_{\bot}}^{R}]&=[\hat{a}_{\omega k_{\bot}}^{R\dagger},\hat{a}_{\omega'k'_{\bot}}^{R\dagger}]=0.
\end{align}


\subsection{The Wightman function}

We evaluate the Wightman function of a massless scalar field in the Minkowski vacuum state along the worldline of an accelerated detector to obtain
\begin{align}\label{Witch}
\mathcal{W}_{\text{acc}}(s)=\frac{\lambda^2}{4\pi}\int_0^\infty d\omega\,\bigg(\cos(\omega s)-i\tanh\left(\frac{\pi\omega}{a}\right)\sin(\omega s)\bigg),
\end{align}
 where we have set $s=\eta-\eta'$, i.e., the correlation function is stationary in the Rindler time.  Note that the spectral density of the environment is  frequency-independent.
 In order to compute Eq. \eqref{Witch}
we have employed the expectation values over the Minkowski vacuum state 
\begin{align}
     \expval{\hat{a}_{\omega k_{\bot}}^{R\dag}\hat{a}_{\omega'k'_{\bot}}^R}&=\frac{1}{e^{2\pi\omega/a}-1}\delta(\omega-\omega')\delta(k_{\bot}-k_{\bot}),\\
     \expval{\hat{a}_{\omega k_{\bot}}^{R}\hat{a}_{\omega'k'_{\bot}}^{R\dag}}&=\frac{e^{2\pi\omega/a}}{e^{2\pi\omega/a}-1}\delta(\omega-\omega')\delta(k_{\bot}-k'_{\bot}),
\end{align}
and used the integral \cite{gradshteyn2014table}
\begin{align}
    \int_0^{\infty}dx \, K^2_{i\mu}(x)=\frac{\pi}{4}\bigg|\Gamma\left(\frac{1}{2}+i\mu\right)\bigg|^2=\frac{\pi^2}{4\cosh(\pi\mu)},
\end{align}
where $\Gamma(z)$ is the gamma function. For an alternative derivation of Eq. \eqref{Witch}, as well as its exact expression see Appendix \ref{app}.

Let us next consider an inertial detector following the worldine $\x (\tau)=(\tau,0,0)$ and interacting with a scalar field 
\begin{align}
    \hat\Phi(t,\xx)=\int\frac{d^2\kk}{\sqrt{(2\pi)^22\omega_{\kk}}}\left(\hat{a}_{\kk}e^{-i(\omega_{\kk}t-\kk\cdot\xx)}+ \text{H.c.}\right),
\end{align}
where $\omega_{\kk}=\sqrt{|\kk|^2+m^2}$, $m$ denotes the mass of the field, $\hat{a}_{\kk}^\dag$ and $\hat{a}_{\kk}$ are the creation and annihilation operators of the filed mode with momentum $\kk$ that satisfy the standard canonical commutation relations. We suppose that the field is at a thermal equilibrium state with temperature $T=1/\beta$. In this case, the Wightman function of the field pulled back to the detector's worldline is
\begin{align}\label{Wight:thermal}
    \mathcal{W}_{\text{th}}(s')=\frac{\lambda^2}{4\pi} \int_0^\infty d\omega_{\kk}\,J(|\kk|) \bigg(&\coth\left(\frac{\beta \omega_{\kk}}{2}\right)\cos\left(\omega_{\kk}s'\right) \nn\\&\quad -i\sin\left(\omega_{\kk}s'\right)\bigg),
\end{align}
where $s'=\tau-\tau'$ and $J(|\kk|)=|\kk|/\omega_{\kk}$. The hyperbolic cotangent that appears in the noise kernel suggests that the noise experienced by the detector is thermal. The correlation function in \eqref{Wight:thermal} has the conventional form of one characterizing an environment  at a thermal state with temperature $T$. Such forms (with spectral densities following a power law $J(|\kk|)\sim|\kk|^\nu$) are common in QBM models.

Beyond the (2+1)-dimensional case considered above, it can be shown that an oscillator detector moving with constant acceleration through the Minkowski vacuum behaves exactly in the same way as an inertial one at an Ohmic (in (3+1) dimensions) or a sub-ohmic (in (1+1) dimensions) heat bath at the Unruh temperature $T_U=a/2\pi$, a behavior that is also implied by the similarity  of the dissipation and noise kernels between the two cases \cite{Hu:Louko,Raval:Hu}. This equivalence offers one way to see the Unruh effect. However, as pointed in \cite{Takagi}, the two pictures: (i) an inertial detector immersed in a thermal field bath at the Unruh temperature and (ii) a uniformly accelerated detector in Minkowski vacuum are not equivalent in the (2+1)-dimensional spacetime.

The relation between the dissipation and noise kernels and the Wightman function through Eq. \eqref{infkern} allows one to evaluate the Fermi golden rule transition rate of an oscillator detector (let's assume it starts out in its ground state) as \cite{Amplif}
\begin{equation}\label{spectrum}
    w(\Omega)=\int^{\infty}_{-\infty}ds\, e^{-i\Omega s}\mathcal{W}(s).
\end{equation}
It is then straightforward to show that in the (2+1)-dimensional case the transition rate of an accelerated detector obeys a Fermi-Dirac distribution $w\sim\text{Exp}^{-1}(2\pi\Omega/a+1)$ at the Unruh temperature, in contrast to the Planckian form $w\sim\text{Exp}^{-1}(2\pi\Omega/a-1)$ observed in  (1+1) and (3+1) dimensions and in the thermal inertial case at all dimensions. The ``statistics inversion" in the distribution that characterizes the detector's transition rate spectrum in odd spacetime dimensions was originally reported in \cite{TakagiInv,Takagi}. Note that this does not mean that the background scalar field bath somehow changed its statistics; each field mode is populated by a Bose factor  \cite{Unruh:odd}.

We notice that neither the discrepancy between the Wightman functions in the accelerated and the thermal inertial cases considered above nor the non-Planckian form of the transition rate of an accelerated detector means that the Unruh effect is not present in (2+1) dimensions. As we shall see next, in the long time limit the detector reaches a thermal equilibrium state at the Unruh temperature. It is the thermality of the detector's asymptotic state that offers a more robust and universal way to interpret the Unruh effect, as has been pointed in \cite{DMCA,DM}.

We next demonstrate that the dependence of the dissipation kernel on the acceleration temperature (see Eq. \eqref{Witch}) gives rise to a temperature dependent damping rate and Lamb shift. This is in contrast to any conventional QBM model where the dissipation kernel (see Eq. \eqref{Wight:thermal}) does not depend on the temperature of the bath and both the dissipation rate and the Lamb shift are temperature independent. 


\subsection{Dissipation and noise}

In order to avoid any convergences in the calculation of the Laplace transform of the dissipation kernel integral we introduce an exponential cut-off in it
\begin{equation}\label{diss2}
    \chi(s)= 
    -\frac{\lambda^2}{4\pi}\int_0^\infty d\omega\, e^{-\epsilon\omega} \tanh\left(\frac{\pi\omega}{a}\right)\sin\left(\omega s\right),
    \end{equation}
where $\epsilon^{-1}$ denotes a high frequency cut-off. Although the regularization of the high frequency behavior by the introduction of the exponential factor $e^{-\epsilon\omega}$ may seems ad hoc, in UDW detector models it is shown that the positive parameter $\epsilon$ can be related to the size of the detector, with the limiting case $\epsilon\to 0^+$ taken after the computation of the integral, to correspond to the case of a point-like detector \cite{Schlicht_2004,Louko:Satz}. Note that also to calculate the Laplace transform of the dissipation kernel through its exact expression contained in \eqref{Wight:app}, a relevant regulator that ensures the existence of the integral transform is needed. 
We then evaluate the Laplace transform of the dissipation kernel integral \eqref{diss2}, for small but finite values of $\epsilon$, to obtain
\begin{align}
    \widehat{\chi}(z)=-\frac{\lambda^2}{4\pi}\left[-\log\left(e^{\text{\textgamma}} \epsilon a\right)-\psi\left(\frac{z}{a}+\frac{1}{2}\right)\right],
\end{align}
where $\text{\textgamma}$ is the Euler-Mascheroni constant and $\psi(z)$ is the psi (digamma) function \cite{NIST}.

Working in the weak-coupling regime we look perturbatively for the poles of the integrand in \eqref{homo:lapl}, i.e., we look for the solutions of equation $z^2+\Omega^2+2\widehat{\chi}(z)=0$. We find that the poles are $z_\pm=-\gamma \pm i\Omega' +\mathcal{O}(\lambda^4)$,
where
\begin{align}\label{diss:rate}
    \gamma=\gamma_0\tanh{\left(\frac{\pi\Omega}{a}\right)}
\end{align}
is the dissipation rate with $\gamma_0=\frac{\lambda^2}{8\Omega}$ denoting the damping constant obtained in the case of inertial motion in a massless field bath at a thermal state,
and $\Omega'^2\equiv\Omega^2+\Omega_R+\delta\Omega$, where
\begin{align}
\Omega_R=\frac{4\gamma_0\Omega}{\pi}\ln\left(e^{\text{\textgamma}} \epsilon \Omega\right)
\end{align}
is an acceleration independent frequency renormalization term and
\begin{align}\label{Lamb:Shift}
    \delta\Omega=\frac{4\gamma_0\Omega}{\pi}\left[\ln(\frac{a}{\Omega})+\text{Re}\psi\left(i\frac{\Omega}{a}+\frac{1}{2}\right)\right]
\end{align}
is a finite frequency shift (the Lamb shift) due to acceleration. To obtain Eqs. \eqref{diss:rate} and \eqref{Lamb:Shift} we have used the functional relation $\text{Im} \psi\left(\frac{1}{2}+ix\right)=\frac{\pi}{2}\tanh(\pi x)$. Note that we have incorporated the divergent shift term $\Omega_R$ into the definition of the oscillator detector's frequency. Alternatively, one may include a second order to the coupling counter-term into the interaction Hamiltonian to compensate
for the renormalization \cite{breuer,Leggett:Review}.

\begin{figure}
    \subfloat[Dissipation rate]{\includegraphics[scale=0.6]{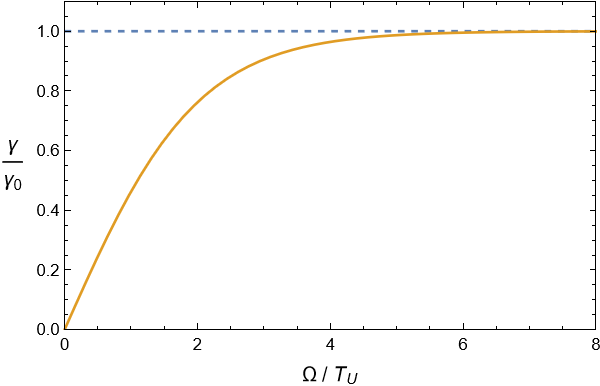}}
    
\subfloat[Frequency shift]{\includegraphics[scale=0.6]{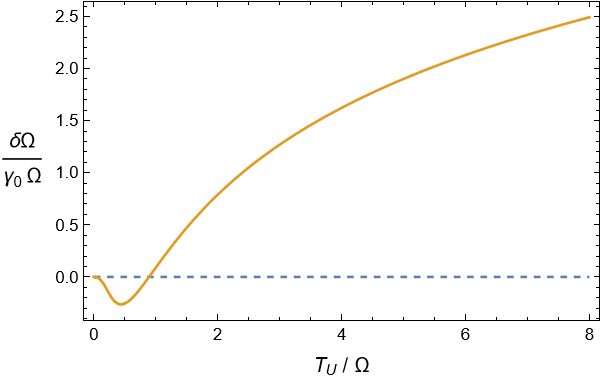}}
    \caption{ Dissipation rate and frequency shift of a uniformly accelerated oscillator detector (solid line) compared to the ones found in the case of an inertial detector immersed in a heat bath at the Unruh temperature $T_U$ (dashed line).}
    \label{figure}
\end{figure}

In Fig. \ref{figure} we plot the acceleration temperature dependent dissipation rate \eqref{diss:rate} and frequency shift \eqref{Lamb:Shift} compared to the ones found in the case of an inertial detector immersed in a heat bath at the Unruh temperature. We observe that the dissipation rate increases with frequency until it reaches the temperature independent value obtained in the thermal inertial case and becomes constant. On the other hand, the frequency shift due to acceleration significantly differs from the one found in the thermal inertial case and which does not depend on the temperature of the field bath.

Finally, employing the Cauchy's residue theorem we evaluate the Bromwich integral \eqref{homo:lapl} to obtain the homogeneous solution 
\begin{align}\label{homo}
    G(\eta)=e^{-\gamma \eta}\frac{\sin(\Omega' \eta)}{\Omega'}.
\end{align}
Note that the psi function $\psi(z)$ is a meromorphic function with simple poles at $z=-n$, $n\in\mathbb{N}_0$. Thus, in the homogeneous solution \eqref{homo} there should be another term with the sum of the residues of the poles of the psi function. However, this extra term, which resembles relevant branch-cut terms in QBM models, gives rise to effects that are significant only at very early times, decaying fast in time \cite{Fleming:Hu:Roura,Charis,TK}. As we are interested in the late-time behavior of the detector we drop this term. For similar non-Markovian poles that appear in the case of a qubit detector and affect its evolution at early times see \cite{DMCA}.

\subsection{Late-time covariances}
In the long-time limit the position correlator in Eq. \eqref{sigma:xx} takes the form 
\begin{align}\label{sxx3}
    \sigma_{11}(\infty)=\frac{2\gamma_0}{\pi}\int_0^{\infty}d\omega\, \widehat{G}(i\omega)\widehat{G}(-i\omega),
\end{align}
where $\widehat{G}(z)$ is the Laplace transform of the homogeneous solution \eqref{homo}. It is given by 
\begin{align}
    \widehat{G}(\pm i\omega)=\frac{1}{\Omega^2+(\gamma\pm i\omega)^2}
\end{align}
and thus Eq. \eqref{sxx3} reads
\begin{align}\label{corpos}
    \sigma_{11}(\infty)=\frac{1}{\pi}\int_0^{\infty}d\omega\, \frac{2\gamma_0\Omega}{(2\gamma_0\Omega)^2\tanh^2(\pi\Omega/a)+(\omega^2-\Omega^2)^2}.
\end{align}
It is also straightforward to obtain the asymptotic form of the momentum correlator $\sigma_{pp}(\infty)$ by noticing that $\dot{\widehat{G}}(i\omega)\dot{\widehat{G}}(-i\omega)=\omega^2\widehat{G}(i\omega)\widehat{G}(-i\omega)$. The remaining covariances $\sigma_{12}(\infty)=\sigma_{21}(\infty)$ vanish. 

In the vanishingly weak coupling limit, $\gamma_0/\Omega\to 0$, the Lorentzian function in \eqref{corpos} is replaced by the delta function
\begin{align}
\delta\left(\frac{\omega^2-\Omega^2}{\Omega^2\tanh(\pi\Omega/a)}\right)
\end{align}
and we obtain
\begin{align}
    \sigma_{11}(\infty)&=\expval{\hat{x}^2(\infty)}=\frac{1}{2\Omega}\coth\left(\frac{\Omega}{2T_U}\right),\\
    \sigma_{22}(\infty)&=\expval{\hat{p}^2(\infty)}=\frac{\Omega}{2}\coth\left(\frac{\Omega}{2T_U}\right),
\end{align}
which describe a thermal state at the Unruh temperature $T_U$ \cite{breuer}.

\subsection{The fluctuation-dissipation theorem}

Taking the Fourier transform $\widetilde{f}(\omega)=\int_{-\infty}^{\infty}ds\,e^{-i\omega s}f(s)$
of the noise kernel in Eq. \eqref{Witch} (or \ref{Wight:app}) we have 
\begin{align}\label{ale}
    \widetilde\nu(\omega)=2\gamma_0\Omega.
\end{align}
Furthermore, the imaginary part of the Fourier transformed dissipation kernel reads
\begin{align}\label{stout}
    \text{Im}\widetilde{\chi}(\omega)=2\gamma_0\Omega\tanh\left(\frac{\pi\omega}{a}\right).
\end{align}
Combining Eqs. \eqref{ale} and \eqref{stout} we obtain the fluctuation-dissipation theorem \begin{align}
    \widetilde\nu(\omega)=\coth\left(\frac{\omega}{2T_U}\right)\text{Im}\widetilde{\chi}(\omega),
    \end{align}
which implies that fluctuations in equilibrium are thermal. It is identical to the conventional form obtained in the case of a system in thermal equilibrium at some temperature $T$ (here this is the Unruh temperature). The relation between the Fourier transforms of the expectation values of the commutator and anti-commutator of the field expressed by the fluctuation-dissipation theorem has also been deduced in \cite{Takagi} from  the form of the detector's spectrum \eqref{spectrum}.

\section{Conclusions}

We studied the response of a uniformly accelerated oscillator detector interacting with a massless scalar field in its vacuum state in the (2+1)-dimensional flat spacetime. We showed that in the weak coupling limit the detector reaches at late times a thermal state at the Unruh temperature. This leads us to suggest that a uniformly accelerated detector in Minkowski vacuum and an inertial one immersed in a thermal field bath at the Unruh temperature behave in the same way no matter what the dimensions of the background spacetime are only in terms of their late time behavior. As we have argued before \cite{DMCA,DM} it is this late time behavior of the detector that offers a robust and universal way to interpret the Unruh effect.

Apart from that, we revised in the framework of open quantum systems an older result by Takagi \cite{Takagi} reporting the non-equivalence between the accelerating and the  thermal inertial case in (2+1) spacetime dimensions. Investigating the implications of this non-equivalence on the evolution of an accelerated oscillator we demonstrated that in the (2+1)-dimensional case both the accelerated detector’s dissipation rate and the shift of its frequency caused by the coupling to the field bath depend on the acceleration temperature, as opposed to analogous QBM models in open systems, where neither of them exhibits temperature dependencies. 

We note that since the Caldeira-Leggett model of Brownian motion and the quantum Langevin approach are generally used to describe the dynamics in many real systems, as, for example, in superconducting circuits elements, it would be interesting to explore the emergence of the unique characteristics of the Unruh (or any Unruh-like \cite{Good}) effect in the (2+1) dimensional spacetime geometry in analogue gravity  experiments \cite{Blencowe,analogue}.

\section{Acknowledgments}
The author wishes to thank Charis Anastopoulos and Theodora Kolioni for fruitful discussions and useful comments during the preparation of
this manuscript. This research is co-financed by Greece and the  European Union (European Social Fund-ESF) through the Operational Programme ``Human Resources Development, Education and Lifelong Learning" in the context of the project ``Reinforcement of Postdoctoral Researchers - 2nd Cycle" (MIS-5033021), implemented by the State Scholarships Foundation (IKY).

\appendix
\section{Alternative derivation and exact expression for the Wightman function Eq. \eqref{Witch}}\label{app}

In the main text, we have employed the Rindler-Fulling quantization scheme \cite{Fulling} for a massless scalar field  in order to obtain the Wightman two-point  function Eq. \eqref{Witch}.
Alternatively, we can calculate the Wightman function in a straightforward manner  through its definition \eqref{infkern},
\begin{align}
    \mathcal{W}(\tau,\tau')&=\int\frac{d^2\kk}{(2\pi)^{2}2|\kk|}e^{-i\kk(t(\tau)-t(\tau')-i\epsilon)}e^{i\kk\cdot(\mathbf{x}(\tau)-\mathbf{x}(\tau'))}\nonumber,\\
\end{align}
along the accelerating detector's trajectory $\x(\tau)=(a^{-1}\sinh(a\tau),a^{-1}\cosh(a\tau),0)$  to obtain
\begin{align}
    \mathcal{W}(\tau-\tau')&=\frac{1}{4\pi}\int_0^{\infty}dke^{-2ika^{-1}\sinh\left(\frac{a}{2}(\tau-\tau'-i\epsilon)\right)}\label{Witch:ex}\\
    &=\frac{1}{4\pi}\frac{1}{2ia^{-1}\sinh\left(\frac{a}{2}(\tau-\tau'-i\epsilon)\right)},
\end{align}
where the limit $\epsilon\to0^+$ is understood in the distributional sense. Note that we have omitted the coupling constant that we previously absorbed into a redefinition of the field in the main text. Furthermore, using the Sokhotsky's formula
\begin{equation}
    \lim_{\epsilon\to0^+}\frac{1}{x\pm i\epsilon}=\mp i\pi\delta(x)+\text{PV}\frac{1}{x},
\end{equation}
where PV denotes the Cauchy principal value, we have
\begin{align}\label{Wight:app}
    \mathcal{W}(s)=\frac{1}{4\pi}\left(\pi\delta(s)-i\frac{\text{PV}}{2a^{-1}\sinh(as/2)}\right)
\end{align}
where $s=\tau-\tau'$, from which we can directly identify the exact expressions for the noise and the dissipation kernels respectively.

We notice that the integral form  \eqref{Witch} of the Wightman function can also be obtained through Eq. \eqref{Witch:ex} by employing the relation \cite{gradshteyn2014table}
\begin{align}
    e^{-i\alpha\sinh(x/2)}&=\frac{4}{\pi}\int_0^\infty d\nu K_{2i\nu}(\alpha)\Big(\cosh(\pi\nu)\cos(\nu x)\nn\\&\quad \quad \quad \quad- i\sinh(\pi\nu)\sin(\nu x)\Big)
\end{align}
and making use of the integral \cite{gradshteyn2014table}
\begin{equation}
    \int_0^{\infty}dx K_{\mu}(x)=\frac{\pi}{2\cosh(\pi\mu/2)}.
\end{equation}
\bibliography{references}
\end{document}